\documentclass[11pt,a4paper]{article}
\usepackage{graphicx}
\usepackage{cite}
\usepackage{latexsym}

\newcommand{\ie}{{\frenchspacing i.e. }}
\newcommand{\beq}{\begin{equation}}
\newcommand{\eeq}{\end{equation}}
\newcommand{\beqa}{\begin{eqnarray}}
\newcommand{\eeqa}{\end{eqnarray}}
\newcommand{\bea}{\begin{eqnarray}}
\newcommand{\eea}{\end{eqnarray}}

\newcommand{\forget}[1]{}

\newcommand{\mbf}[1]{$\mathbf{#1}$}
\newcommand{\assign}{\mbox{:=}~}      % Assignament operator
\newcommand{\var}[1]{\mbox{$\mathit{#1}$}}      % Style of variables
\newcommand{\const}[1]{\mbox{$\mathsf{#1}$}}    % Style of constants
\newcommand{\func}[1]{\mbox{{\scshape #1}}}

\newcommand{\s}{\sigma}
\newcommand{\set}[1]{\{#1\}}

\newcommand {\fig}[3]{\resizebox{#2}{!}{\rotatebox{#3}{\includegraphics{#1}}}}

\begin{document}
\begin{titlepage}

\begin{flushright}
LU TP 01-23\\
July 2, 2001
\end{flushright}

\vspace{.25in}

\LARGE

\begin{center}
{\bf Stability of the  Kauffman Model}\\
\vspace{.3in}
\large

Sven Bilke and   
Fredrik Sjunnesson\footnote{\{sven,fredriks\}@thep.lu.se}\\ 
\vspace{0.10in}
Complex Systems Division, Department of Theoretical Physics\\ 
University of Lund,  S\"{o}lvegatan 14A,  S-223 62 Lund, Sweden \\
{\tt http://www.thep.lu.se/complex/}

\end{center}
\vspace{0.25in}

\large
{\bf Abstract:}

Random Boolean networks, the Kauffman model, are revisited by
means of a novel decimation algorithm, which reduces the networks
to their dynamical cores.  The average size
of the removed part, the stable core, grows approximately 
linearly with  $N$, the number of nodes in the original networks. We show 
that this can be understood as the percolation of the stability signal 
in the network. The stability of the dynamical core is investigated 
and it is shown that this core lacks the well known stability observed in full Kauffman 
networks. We conclude that, somewhat counter-intuitive,  the remarkable stability 
of Kauffman networks is generated by the dynamics of the stable core. 
The  decimation method  is also used  to simulate large critical 
Kauffman networks. For networks up to $N=32$ we perform 
{\em full enumeration}  studies. Strong evidence is provided for that 
the number of limit cycles grows linearly with $N$. This result is in sharp
contrast to the often cited $\sqrt{N}$ behavior.  

\begin{center}
Submitted to {\it Physical Review} {\bf E}
\end{center}

\large

\vspace{0.8in}


\end{titlepage}

%\Large
\large

\newpage

%---------------------------
\section{Introduction}
%---------------------------
Boolean networks were introduced by Kauffman \cite{Kauffman:69,Kauffman:93} 
as simplified models of the complex interaction in the regulatory 
networks of living cells. The binary variable $\sigma _i$  encodes the
activity of the effective ``gene'' $i$; expressed or not expressed.
Depending upon the initial state,
the system evolves to one of the several limit cycles.
In the biological picture, the
different limit cycles are interpreted as different cell types.
One of Kauffman's motivations for investigating 
these networks was the idea that the structure of genetic networks present in
nature is not only determined by selection. Rather, 
a good fraction of the network functionality is inherent in the ensemble 
of regulatory networks as such. In fact, he found an ensemble
of critical Boolean networks ``on the edge of chaos'' that captures some
features observed in nature. These Boolean networks show a remarkable 
stability; in most cases small perturbations in the state of the network do
not change the trajectory to a different limit cycle. This is desirable 
in the biological interpretation since stability of genetic regulatory networks
against small fluctuations is a crucial property.
Another striking observation is that the number of limit cycles 
for the critical  Boolean networks grows as a square-root of the system size 
\cite{Kauffman:69, Kauffman:93}.
This is an analogy to 
multicellular organisms, where it is found empirically that 
the number of cell-types also grows approximately as the square-root of
the genome-size.

The model also exhibits analogies \cite{DF86} with infinite range spin glasses
\cite{MPV:87}. In the framework of an annealed approximation  \cite{DP86}, some
of the previous numerical observations concerning a phase transition  between a
{\em frozen} and a {\em chaotic} phase in the model could be understood. The
average  limit Hamming distance $d_h$, the number of bit-wise differences
between two random configurations, was used as an order parameter. In the
frozen phase one has $d_h = 0$ for infinite systems whereas in the chaotic
phase one has $d_h \neq 0$. The parameter driving the  transition is the 
probability $p$ that two different inputs to a Boolean variable $\sigma_i$ 
give raise to different values.
\forget{The critical coupling is at $p = p_c(K)$ with 
$2 p_c ( 1 - p_c) = 1 / K$, where $K$ is the number of input variables 
to $\sigma_i$.} 
In \cite{BP95} the annealed
approximation was extended to provide distributions for the number and the
length of the limit cycles. Also, good agreement between the results from
the annealed approximation and the numerical calculations  was demonstrated in
the chaotic phase. An alternative order parameter $s$, the fraction of variables that
are stable, \ie evolve to the same fixed state independently of the initial state,
was introduced in \cite{Flyvbjerg:88}. In the infinite size limit one has $s=1$ in the frozen phase,
whereas $s\neq1$ in the chaotic phase.
In \cite{Parisi:98} the concept of relevant variables was introduced.
A variable $\sigma_i$ is {\em not} relevant, if it is stable and/or no variable state
depends on $\sigma _i$. The relevant variables are of interest since they
contain all information about the asymptotic dynamics of the network, \ie the number of
limit cycles and their cycle lengths.\forget{ With approximate
arguments they predicted that their number scales as $\sqrt{N}$ on the critical line,
while it is linear with N in the chaotic phase and independent of N in the frozen phase.}

In this work we focus on the stability of the Kauffman model and how this property is
related to the stable core of the network. The probability that inversion of a single
variable will make the system end up in a different limit cycle is known to be small and 
approaches zero for large networks. However, we find that if the network is reduced to its relevant
core, this probability in drastically raised and increases slightly with the system size.

To facilitate this study we introduce a method that removes
variables that {\em cannot} be relevant by inspection of 
transition functions and  network connectivity. The resulting reduced network contains 
{\em all} relevant variables and possibly some irrelevant ones. Since all relevant variables
are included it will have exactly the same asymptotic dynamics as the original network
even though the total number of variables is drastically reduced.
We find that resulting number of variables is close to the true number of relevant variables. This 
indicates that properties of the stable core can mostly be understood by the comparatively
trivial interactions detected in the decimation procedure. 

The decimation procedure can also be used to reduce the bias in the estimate of some 
observables like for example the number $n_c$ of limit cycles. Different from earlier works
we do not observe a $\sqrt{N}$ scaling, but rather a linear growth of $n_c$ with the system 
size.    

%---------------------------
\section{The Kauffman Model}\label{kn}
%---------------------------
A random Boolean  network essentially is a cellular automaton with $N$ binary 
state variables $\sigma _i$.  These  evolve synchronously according to the
transition functions $f_i(\{\sigma\})$, which are  chosen randomly at time
$t=0$ and are then kept fixed. In the Kauffman model $f_i$ are constrained 
to depend on at most $K$ different randomly chosen input variables: 
\beq 
        \sigma_i(t) = f_i[\sigma_{v_i^1}(t-1), \ldots, \sigma_{v_i^K}(t-1)],
\label{upd}
\eeq
for every variable $\sigma_i$. The integers $\set{v_i^1,\ldots,v_i^K}$ define the input
connections to variable~$i$.

The transition function $f_i$ maps each possible combination of input signals to  
Boolean output values. These output values are independently set to 
$true$ or $false$ with probabilities $p$ and $1-p$ respectively.
This makes some functions
independent of some or all of its $K$ input variables.
Furthermore, depending on $K$ and $p$, a finite fraction of the state 
variables $\sigma _i$ are not used by any of the transition functions.

The random Boolean network is a deterministic system. Given
the state variables at some time, the future trajectory of the $\sigma _i$ is
known. The volume of the state space is finite, therefore all trajectories must 
posses a limit cycle. Besides the stability of the system the number of limit cycles,
the length distribution of the cycles 
and transient trajectories are well established observables for this model. In 
numerical simulations it is in general not possible to probe the models whole 
state space, except for very small systems. The volume of the $\{\sigma \}$ state
space grows exponentially and the number of graphs $\{ f_i \}$ even super-exponentially. 
The commonly used strategy for exploring this model therefore contains two approximations.
\begin{enumerate}
\item A small fraction of all possible networks is used as a representative ensemble.
\item For each network only a subset of the state space is probed.
\end{enumerate}
Point 2 introduces a systematic bias to the number of limit cycles since not
all of them will be found. In the results section 
we will re-analyze the number of cycles after decimation of
irrelevant nodes. This allows {\em full enumeration} of state space for up to
$N=32$. In this way we get an improved estimate for the scaling of the 
respective observables with the system size.

%---------------------------
\section{The Decimation Procedure}
%---------------------------
It is well known that some variables in a Kauffman network evolve to the same steady
state independently of the initial configuration. 
These {\em stable} variables are clearly irrelevant for the asymptotic behavior of the network.
The same holds for those variables that do not regulate any other variable,
\ie no transition functions is dependent on them.
In fact, as pointed out in \cite{Parisi:98}, for a variable to be relevant
it has to be unstable and regulate some unstable variables that in turn regulate others
and so on.
In other words, a variable is relevant if and only if it is unstable and regulates other
relevant variables.

By definition, in the frozen phase the fraction of stable variables goes to unity as
$N$ goes to infinity. Therefore, a large fraction of the variables are likely to be stable
even for finite $N$. Since the irrelevant variables includes all stable
variables, a considerable part of a network does not affect the asymptotic
dynamics at all.
The process of identifying the irrelevant variables can be divided into two separate steps.
Firstly, the stable variables are identified. 
Secondly, the variables that do not regulate any unstable variables are identified.

Identifying stable variables is in principal easy, but
computationally demanding. In \cite{Parisi:98} this was done by  performing simulations of the dynamics
of the system and monitoring which variables were in the same state in all probed
limit cycles. However, finding all limit cycles essentially means that all $2^N$ possible
states have to be probed, which is possible only for very small networks.
Since a variable that is stable within the probed limit cycles may change state within some of the
unprobed limit cycles, searching a fraction of state space will in some cases overestimate number
of stable variables. \forget{However, because the variables that are identified as unstable are
guaranteed to be unstable, the set of variables identified as stable is a super-set
of the true answer.}

Here we introduce an alternative method, which by pure inspection
of the connectivity and the transition functions
of a network identifies variables that {\em must} be stable. \forget{We hereby identify 
a subset of all stable variables.}
The basis for our approach is that transition functions dependent on no input variables 
give a constant output, \ie the corresponding variable is stable.

As stated above, some transition functions are independent of all their
input variables, \ie they are constants. This means that the corresponding
variables will be stable (after the initial time step) and
a transition function that is dependent on such variable will
receive a constant signal. By replacing the {\em stable} input variable with the
corresponding {\em constant} value, the number of input variables is reduced. For each
replaced variable the input state space is reduced by a factor $1/2$
and within this subspace the rule may be independent
of yet other input variables. If in the end even this rule become a constant, the 
corresponding variable is stable (after a transient time), and can be replaced by a
constant. Therefore, we have to repeat this procedure until no more stable variables
are found. We summerize the method as follows:
\begin{enumerate}
\item For every updating function, $f_i$, remove all inputs it does not depend upon.
\item For those $f_i$ with no inputs, clamp the variable $\sigma_i$ to the corresponding constant value.
\item For every $f_i$, replace the clamped inputs with the corresponding constant.
\item If any variable has been clamped, repeat from step 1.
\end{enumerate}
For a pseudo-code  description of this method see Appendix~(\ref{pseudo}).

\begin{figure}[h]
\centerline{\fig{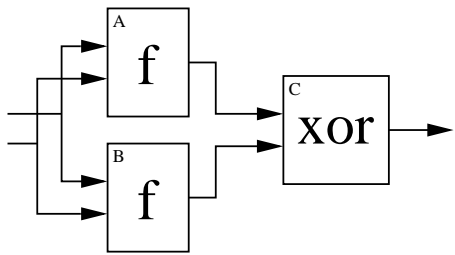}{6cm}{0}}
\caption{$A$ and $B$ have arbitrary but identical updating functions and $C$ implements $xor$.
Since $A$ and $B$ also have the same inputs their outputs will be identical. Thus, $C$ will
always output $false$, i.e.~$C$ is stable and can be removed. $A$ and $B$ are now non-regulating
and can be removed too.}
\label{hardNet}
\end{figure}

It is clear that our method sometimes does not find all stable variables. We see an example of
such a situation in Fig.~\ref{hardNet}. Here the inputs to a function are coupled
logically and hereby confined to a subspace of possibilities. Within this subspace
the otherwise unstable variable is stable. The figure illustrates just one
of the possible couplings between inputs. 

Once the stable variables are identified and removed from the network the non-regulating variables 
can be removed iteratively. Since our method keeps all relevant variables the resulting network
will have exactly the same asymptotic dynamics as the original network.

\forget{
If the stable variables were correctly
identified the resulting network would contain just the relevant variables. Since our method
identifies just a subset of the stable variables the resulting network will contain all
relevant variables along with some irrelevant variables. It is important to notice
that by keeping some irrelevant variables we do not change the asymptotic dynamics
of the system, \ie the number of cycles and their cycles lengths are the same as the original
network. The method used in \cite{Parisi:98} identifies a super-set of the
truly stable variables. Hereby, the resulting network will contain no irrelevant variables,
but may lack a few relevant variables. This may change the asymptotic dynamics of the system.
}
%---------------------------
\section{Results}
\begin{figure}[h]
\centerline{\fig{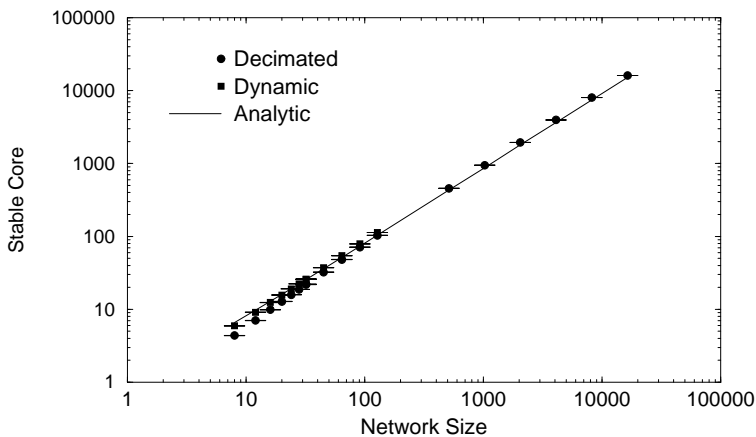}{10cm}{-90}}
\caption{The size of the stable core -- the number of variables going to the same constant 
         value independently of the start-configuration -- as a function of the network 
         size. The size of the relevant core estimated with our decimation procedure
	 (circle), the observation of the dynamics \cite{Parisi:98} and 
	Eq.~(\ref{stable1}) and (\ref{stable2}) are in very good agreement.}
\label{effective}
\end{figure}

%---------------------------
Let us start by analyzing the size of the stable core  
as a function of the system size $N$.  In  Fig.~\ref{effective} the 
the  size of the stable core $N^*$, identified by the decimation procedure described above,
is shown. Each data point is averaged over $10^4$ instances of networks.  For comparison 
the size of the stable core $N^+$ as estimated by the method used in
\cite{Parisi:98} is also plotted. The latter procedure is based on observations of the dynamics 
of the full network and identification of nodes acquiring the same constant value independently of the
start-configuration. Since only a small part of the state space can   
be probed in practice, the number $N^+$ is biased to overestimate  the true size $\eta$ of the
stable core. On the other hand, our decimation procedure underestimates $\eta$ because
some configurations, like the one depicted in  Fig.~\ref{hardNet},  which 
may lead to stable variables are not identified. Therefore, we have 
$N^+ \leq \eta \leq N^*$. 

It is somewhat surprising to observe $N^+ \approx N^*$, which 
indicates that properties of the stable core can mostly be understood by the comparatively
trivial interactions detected in the decimation procedure. The probability $s$ for a node to belong
to the stable core can be estimated by using Eq.~(2) in \cite{Flyvbjerg:88}
\beq
 s(t + 1) = \sum _{k=0}^K s(t) ^{K-k} (1 - s(t))^k {K \choose k} p_k,
 \label{stable1}
\eeq
where $p_k$ is the probability that a transition function for given values of $K-k$ of its input
variables is independent of its other $k$ inputs.
This equation describes the growth of the stable core with the time.
At $t=0$ only nodes which happen to have a constant transition
function are stable. At later times non-constant transition functions, which receive 
inputs from stable nodes, can acquire a constant value.
In \cite{Flyvbjerg:88} Eq.~(\ref{stable1}) was used as a self-consistency 
equation for infinite systems, \ie letting $t \to \infty$. In a finite system,
the iteration has to stop at some time $T$, which reflects a characteristic length
in the network, the maximal distance a signal can flow before it reaches
all nodes. The length scale is set by the average distance (in number of links) 
a signal can travel. The signal pathway in a sparse directed random graph with only
a few loops is approximately a branched polymer, where it is known (see e.g. \cite{amb97})
that the average distance grows polynomial, \ie $ T \sim  c N ^{\gamma}$. We have fitted 
the constants $c$ and $\gamma$ numerically to our data and find $\gamma = 0.32(3)$. 

After removing the $s(T)$ stable nodes, the decimation procedure also eliminates
the leaves in the interaction, \ie those non-constant vertices with out-degree $q(t=0) =0$. 
This changes the out-degree of the remaining nodes. Therefore, this procedure is repeated 
until no more nodes with $q(t)=0$ are found. The fraction $P_l$ of (indirect) leaves can be 
estimated by the self consistent equation  
\beq
P_l  = 
  \sum_{q=1}^{\infty} P(q | \tilde{N}, \tilde{K})  P_l^q,
\label{stable2}
\eeq
where $\tilde{N} = N(1 - s)$ is the number of nodes after removing the $\eta$ constant nodes,
$\tilde{K}$ is the average in-degree, an $P(q | \tilde{N}, \tilde{K})$ the distribution 
of the out-degree $q$ given in Appendix~B.
We solved  Eq.~(\ref{stable1}) and (\ref{stable2}) numerically, the resulting graph 
is also shown in Fig.~\ref{effective}, which  is in very good agreement 
with our numerical results.

\begin{figure}[h]
\centerline{\fig{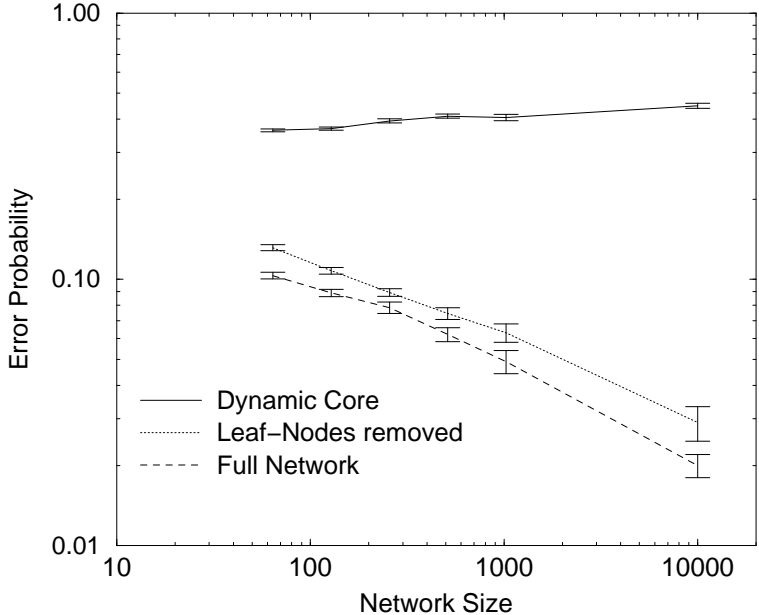}{10cm}{-90}}
\caption{The probability to be pushed out from a limit cycle by the inversion of a randomly
         chosen variable. For the (full) Kauffman-network (dashed line) the error probability 
         scales to zero for large lattices. This behavior is not changed if one does only
         a geometric reduction of leaf-variables (dotted line). For the dynamical core (full line) 
         the tolerance  against small fluctuations in the state space is completely lost.   
}
\label{stability_pic}
\end{figure}

One of the important features of Kauffman's model is the intrinsic stability of  critical 
Boolean networks. How does decimation affect this behavior? While the network reduction does 
not change the number and the length of limit cycles, the size of the basins of attraction 
has to be reduced because the state space is shrunken by orders of magnitude. 
To get a quantitative picture, we analyze the network stability with respect to the inversion 
of one randomly chosen variable, after the state trajectory has reached a limit cycle. 
The probability to end up in a {\em different} limit cycle compared to the undisturbed system is 
shown in Fig.~\ref{stability_pic}. If a limit cycle has not been found within $10^5$ steps the
network is discarded. For the full network we observe the well known stability, the probability 
to end up in a a different limit cycle asymptotically approaches zero for large lattices.
By contrast, for the 
decimated network the error-probability grows slowly with  the network size and
the stability is essentially lost. This means, the tolerance against perturbations observed
for Kauffman networks is mostly generated by the stable core: in most cases the perturbed signal
is ``lost'' in the stable core and the full network remains unaffected. It has recently been 
argued \cite{barabasi_protein} that the in-homogeneous, for example scale-free, {\em  geometry}
of real world networks is underlying the stability of these systems. 
Here we find the opposite: stability is primarily generated dynamically by the propagation 
(Eq.~\ref{stable1}) of the stable core in the network {\em logic}.  The homogeneous geometry 
plays only a secondary role: if just a geometric reduction of the network is performed,
\ie the leaf-variables are removed (see the discussion of of Eq.~(\ref{stable2})),
the error-sensitivity is almost unchanged compared to the full network.  

\begin{figure}[h]
\centerline{\fig{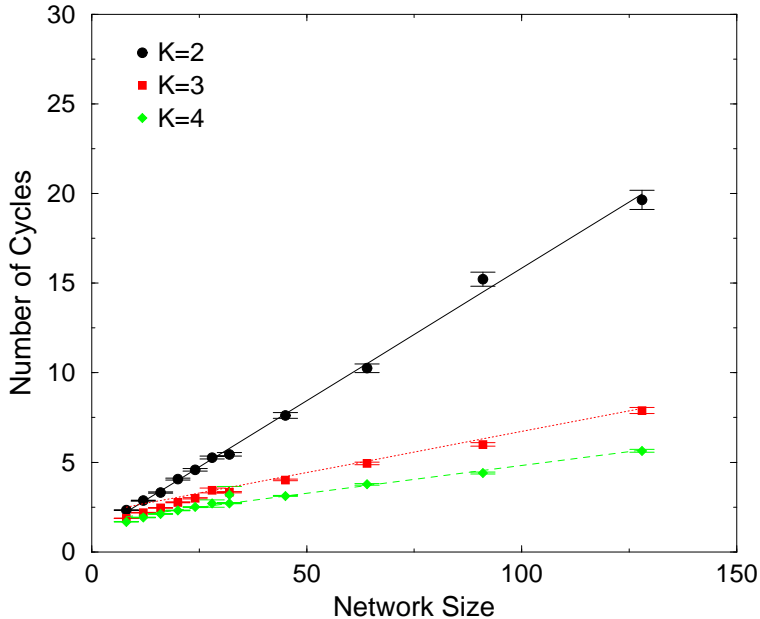}{10cm}{-90}}
\caption{The number of limit cycles as a function of the network size for critical Boolean
         networks with $K=2,3,4$ inputs. The lines connection the data points are {\em linear}
         interpolations with $\chi ^2 / \mbox{D.O.F} = 1.5~(K=2), 1.6~(K=3), 1.4~(K=4)$.
	 A $\sqrt{N}$ behavior fits the data much worse with    
         $\chi ^2 / \mbox{D.O.F} = 11.5~(K=2), 8.6~(K=3), 8.5~(K=4)$.
}
\label{scalnc}
\end{figure}
The decimation of constant variables from the network enables us to probe a much larger fraction
of the state space for a given network. Therefore one may expect to get an better estimate
for the number of limit cycles $n_c$, which with the commonly used method tends to be underestimated, 
because some limit cycles may have been missed due to the huge state space.  By decimating the networks
we can fully enumerate the state space for $N\leq 32$ and hereby get an unbiased estimate. For larger
systems we use the standard method with $1000$ restarts on each of the reduced networks.
Not unexpectedly we observe a small discontinuity in the curve at the point were the simulation scheme is changed.
In Fig.~\ref{scalnc} we plot $n_c$ as a function of the network size $N$.
We do not find the quite often cited
$\sqrt{N}$ behavior for this observable. Rather, we find a linear growth with $N$.
A possible explanation for the  different results obtained in some earlier works may be 
the bias introduced by the standard method in combination with lower computational power.

%---------------------------
\section{Summary}
%---------------------------
The source of the remarkable error tolerance of critical Kauffman model is
identified as the ``dynamics of the stable core''. While this seems to be a 
contradiction in terms, it quite nicely describes the percolation-like process,
which underlies the propagation of the ``stability'' signal. Starting from 
the relatively few nodes with transition functions which do not at all depend
on their inputs, the islands of frozen states grow in time by the interaction with 
the already stable nodes. This process is only limited by the finite size of 
the system.  A small  fluctuation in the state of the system will most probably
not propagate through  the stable core and therefore in most cases has no effect. 
We demonstrate this by reducing given random networks to the dynamical core, where 
most of the stable, irrelevant variables have been removed. The stability against small 
fluctuations for these networks is reduced by orders of magnitude and will probably go 
to zero for infinite networks. It is interesting to observe that these effects are
mostly driven by the network {\em logic} and not by the network geometry. 

For the identification of the dynamical core we have developed a decimation procedure, 
which is based on inspection of the networks connectivity and logic. The 
relatively simple procedure works surprisingly well. The results for the size of the 
stable core are in very good agreement with the values obtained by observing the
dynamics of state-space trajectories in the full network \cite{Parisi:98}.  

As a by-product we use the reduced networks to get an improved estimate for the number of 
limit cycles as a function of the network size. 
\forget{Our results almost surely rule out}
We find that the number of limit cycles grows linearly with $N$, which is in sharp contrast to
the square-root behavior reported by other groups. Even though this $\sqrt{N}$ behavior
was an interesting analogy with multi-cellular organisms (with approximately $\sqrt{N}$
 different cell types for genomes with genome size $N$), our result does in no way reduce
the importance of Kauffman networks as an example of self-organized order.

{\bf Acknowledgments:}
We have benefitted from discussions with C. Peterson. 
This work was in part supported by the Swedish Foundation for Strategic 
Research and the Knut and Alice Wallenberg Foundation through the SWEGENE 
consortium.

\appendix
\renewcommand{\theequation}{\Alph{section}.\arabic{equation}}    

\renewcommand{\thesection}{\Alph{section}}
\renewcommand{\thesubsection}{\arabic{subsection}}
\renewcommand{\thesubsubsection}{\alph{subsubsection}}
%\renewcommand{\theequation@prefix}{\thesection}
%\@addtoreset{equation}{section}
%\addcontentsline{toc}{section}{\protect\numberline{APPENDIXES\hskip
%0pt plus1fill minus1fill\relax}{}}     

%---------------------------
\section{Pseudo Code of the Decimation Procedure}\label{pseudo}
%---------------------------

We remind ourselves of the definition of the Kauffman model.
$\s_i(t)$ denotes the state of variable $i$ at time $t$ which is determined by
$\sigma_i(t) = f_i[\sigma_{v_i^1}(t-1), \ldots, \sigma_{v_i^K}(t-1)]$
with $\set{v}_i = \set{v_i^1,\ldots,v_i^K}$ defining the connectivity of the network.

We represent the network with a number of nodes, each containing an updating function $f_i$
and the list of nodes sending it signals $\set{v}_i$. The function $\func{decimate}$ reduces
the network $Net$ to its relevant core by first removing the stable variables and then those
variables that do not regulate any variable.

\begin{center}
\fbox{\parbox{\textwidth}{\small{
\begin{tabbing}
\ \ \= \kill
\>$node_i=[f_i, \set{v}_i]$\\
\>$Net=\set{node_1,\dots,node_N}$\\[.5cm]

\>{\large \func{decimate}}(\var{Net})\\
\>	\mbf{do} \= \\
\>	         \> \var{oldNet} \assign \var{Net}\\
\>	         \> \mbf{for} \= each \var{n} $\in$ \var{Net} \mbf{do}\\
\>	         \>           \> \func{remove\_unused\_inputs}(\var{n}, \var{Net})\\
\>	         \>           \> \func{remove\_if\_no\_inputs}(\var{n}, \var{Net})\\
\>        \mbf{while} \var{oldNet} $\neq$ \var{Net}\\
\>        \mbf{do} \= \\
\>	         \> \var{oldNet} \assign \var{Net}\\
\>	         \> \mbf{for} \= each \var{n} $\in$ \var{Net} \mbf{do}\\
\>		 \>           \> \func{remove\_if\_no\_outputs}(\var{n}, \var{Net})\\
\>        \mbf{while} \var{oldNet} $\neq$ \var{Net}\\[.5cm]

\>{\normalsize \func{remove\_unused\_inputs}}(\var{n}, \var{Net})\\
\>        \mbf{for} \= each \var{i} $\in$ \var{\set{v}_n} \mbf{do}\\
\>                  \> \var{flag} \assign \const{true}\\
\>                  \> \mbf{for} \= each configuration of inputs \mbf{do} \\
\>                  \>           \> \mbf{if} \= flipping \var{\s_i} changes outcome of $f_n$ \mbf{then}\ \ \ \\
\>	          \>           \>          \> \var{flag} \assign \const{false} \\
\>                  \>  \mbf{if} \= \var{flag}=\const{true} \mbf{then}\\
\>	          \>           \> replace \var{\s_i} with constant value \const{true} in $f_n$\\
\>	          \>           \>remove \var{i} from \var{\set{v}_n}\\[.2cm]

\>{\normalsize \func{remove\_if\_no\_inputs}}(\var{n}, \var{Net})\\
\>	\mbf{if} \= \var{\set{v}_n}=$\emptyset$ \mbf{then}\\
\>   	         \> \mbf{for} \= each \var{m} with \var{n} $\in$ \var{\set{v}_m} \mbf{do}\\
\>	         \>           \> replace \var{\s_n} with constant value $f_n(\emptyset)$ in $f_m$\\
\>	         \>           \> remove \var{n} from \var{\set{v}_m}\\[.2cm]

\>{\normalsize \func{remove\_if\_no\_outputs}}(\var{n}, \var{Net})\\
\>	\mbf{if} \= \var{n} $\notin$ \var{\set{v}_m} for every \var{m} \mbf{then}\\	
\>   	         \> remove \var{n} from \var{Net}
\end{tabbing}
}}}
\end{center}

%---------------------------
\section{In- and Out-degree distribution}
%---------------------------
The reduced number $\tilde{K}$ of inputs after the decimation described in Eq.~(\ref{stable1}) is the 
expectation value of the in-degree for the number of inputs from a non-stable variable: 
\beq
  \tilde{K} =  \frac{ 1 (1 - s(T))  + 2  (1 - s(T))^2 } {1 - s(T) } = 1 + 2 ( 1 -s(T) ).
\eeq

The out-degree distribution for a node in the  random network can be understood by enumerating the 
number of ways the $ N K$ links can be distributed over this node and the $N-1$ remaining nodes,
weighted by the corresponding probabilities to choose the nodes:  
\beq
 P(q | N, K) = \left ( \frac{1}{N} \right )^q 
                               \left ( \frac{N - 1}{N} \right )^{N K - q}
                               {{N K} \choose q}.
\eeq

%\addcontentsline{toc}{section}{References}


\begin{thebibliography}{99}

\bibitem{Kauffman:69}
        S. A. Kauffman, J. Theor. Biol. {\bf 22}, 437 (1969).

\bibitem{Kauffman:93}
        S. A. Kauffman, \emph{The Origins of Order} (Oxford University Press, 1993).

\bibitem{DF86}   B. Derrida and H. Flyvbjerg, J. Phys. A: Math. Gen. {\bf 19}, L1003 (1986).

\bibitem{MPV:87} M. Mezard, G. Parisi and M. A. Virasoro,
	{\em Spin Glass Theory and Beyond} (World Scientific, Singapore, 1987).

\bibitem{DP86}  B. Derrida and Y. Pomeau, Biophys. Lett. {\bf 1}, 45 (1986).


\bibitem{BP95}  U. Bastolla and G. Parisi, Physica D {\bf 98}, 1 (1996).

\bibitem{Flyvbjerg:88}
        H. Flyvbjerg, J. Phys. A: Math. Gen. {\bf 21}, L955 (1988).

\bibitem{Parisi:98}
        U. Bastolla and G. Parisi, Physica D {\bf 115}, 203 (1998).

\bibitem{amb97} J. Ambj{\o}rn, B. Durhuus, T. Jonsson,  
                \emph{Quantum Geometry} (Cambridge University Press, 1997).

\bibitem{barabasi_protein} H. Jeong, S. P. Mason, Z. N. Oltvai, 
A.-L. Barabasi, Nature {\bf 407}, 651 (2000).


 

\end{thebibliography}
\end{document}